\documentclass[journal]{IEEEtran}
\IEEEoverridecommandlockouts
\usepackage{cite}
\usepackage{amsmath,amssymb,amsfonts}
\usepackage{graphicx}
\usepackage{xcolor}
\usepackage{amsmath}

\usepackage{enumitem}
\usepackage{multirow}
\definecolor{Gray}{gray}{0.9} 
\definecolor{LightBlue}{RGB}{173, 216, 230} 
\usepackage{textcomp}
\usepackage{xcolor}
\usepackage{graphicx} 
\usepackage{booktabs} 
\usepackage{amsmath}
\usepackage{amssymb}
\usepackage{algorithm}
\usepackage{algorithmic}
\usepackage{subcaption}
\begin{document}

\title{CITADEL: Continual Anomaly Detection for Enhanced Learning in IoT Intrusion Detection}

\author{
Elvin Li$^{*}$,
Onat~Gungor$^{*}$,
Zhengli~Shang,
and~Tajana~Rosing
\thanks{* Elvin Li and Onat Gungor contributed equally to this work.}
\thanks{Department of Computer Science and Engineering, University of California, San Diego, CA 92093 USA (email: \{ell009, ogungor, z4shang, tajana\}@ucsd.edu).}
}

\maketitle

\begin{abstract}
The Internet of Things (IoT), with its high degree of interconnectivity and limited computational resources, is particularly vulnerable to a wide range of cyber threats. Intrusion detection systems (IDS) have been extensively studied to enhance IoT security, and machine learning-based IDS (ML-IDS) show considerable promise for detecting malicious activity. However, their effectiveness is often constrained by poor adaptability to emerging threats and the issue of catastrophic forgetting during continuous learning. To address these challenges, we propose CITADEL, a self-supervised continual learning framework designed to extract robust representations from benign data while preserving long-term knowledge through optimized memory consolidation mechanisms. CITADEL integrates a tabular-to-image transformation module, a memory-aware masked autoencoder for self-supervised representation learning, and a novelty detection component capable of identifying anomalies without dependence on labeled attack data. Our design enables the system to incrementally adapt to emerging behaviors while retaining its ability to detect previously observed threats. Experiments on multiple intrusion datasets demonstrate that CITADEL achieves up to a 72.9\% improvement over the VAE-based lifelong anomaly detector (VLAD) in key detection and retention metrics, highlighting its effectiveness in dynamic IoT environments.
\end{abstract}


\section{Introduction}\label{sec:intro}

\IEEEPARstart{T}he rapid proliferation of the Internet of Things (IoT) devices has transformed sectors such as healthcare, manufacturing, smart homes, and urban infrastructure~\cite{chataut2023unleashing}. As IoT networks become increasingly pervasive, their interconnected and resource-constrained nature introduces unique security challenges~\cite{cook2023security}. The exponential growth in the number and heterogeneity of IoT devices has significantly expanded the attack surface, rendering these networks increasingly attractive targets for cyber threats. Recent studies report a sustained rise in IoT-focused attacks, including botnets, spoofing, and denial-of-service, which exploit weak authentication and insecure firmware updates~\cite{aouedi2024survey}. These security challenges underscore the critical need for advanced and context-aware protection mechanisms, with a particular emphasis on Intrusion Detection Systems (IDS) specifically engineered to accommodate the unique operational constraints and architectural characteristics of IoT environments. Machine Learning (ML) has become a pivotal approach for IoT-based IDS, owing to its ability to autonomously learn patterns from historical system data, thereby reducing the reliance on extensive manual feature engineering \cite{gungor2024rigorous,gungor2024roldef}. Despite their promising capabilities, state-of-the-art ML-IDS solutions face two challenges: (i) the limited availability of labeled attack data for training and (ii) the difficulty in adapting to the dynamic and continuously evolving nature of both benign and malicious network behaviors.

\begin{figure}
    \centering
    \includegraphics[width=\linewidth]{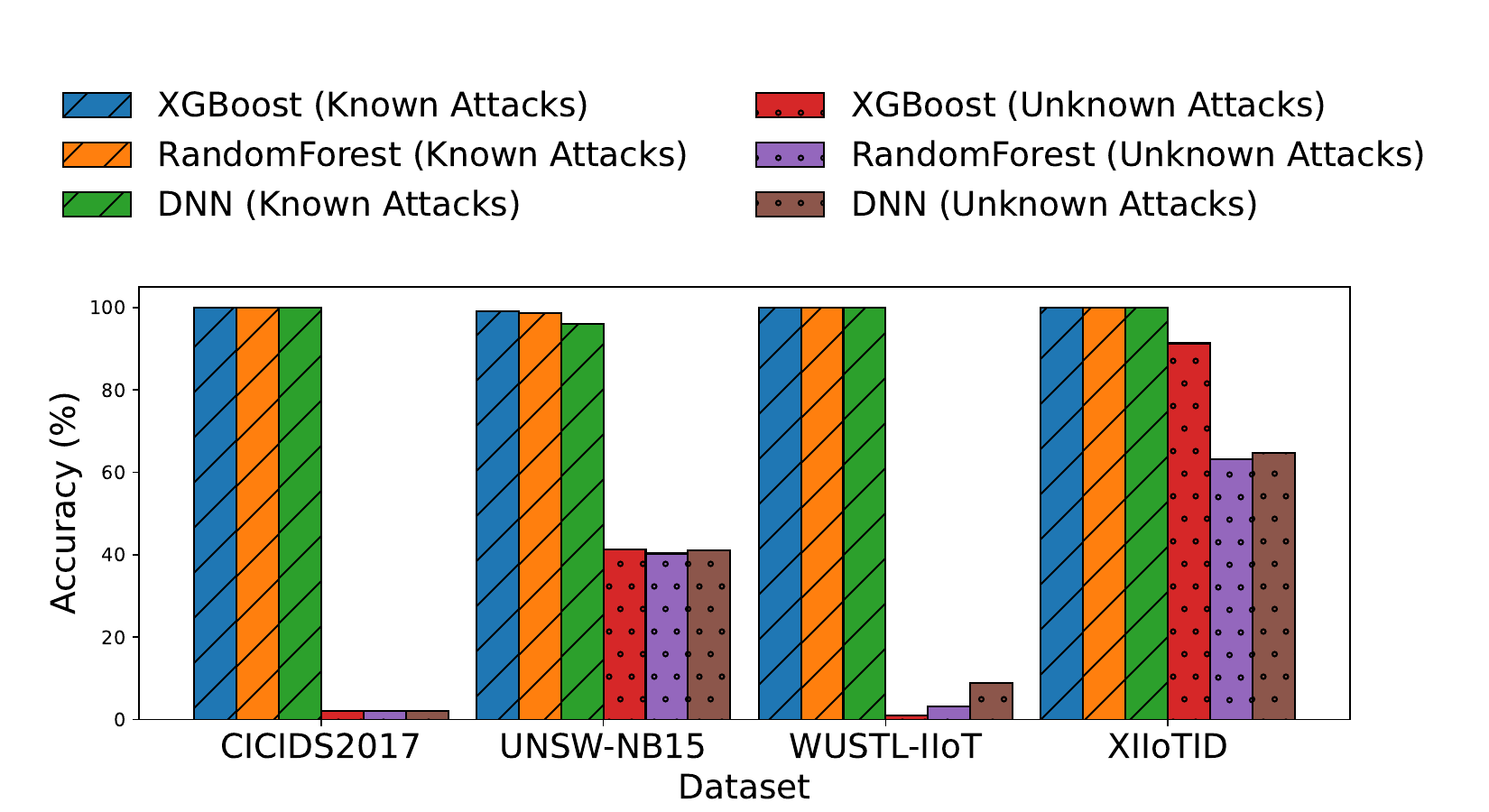}
    \caption{State-of-the-art supervised ML intrusion detection performance on known and unknown attacks}
    \label{fig:unknown_attacks}
\end{figure}

The first challenge is pronounced in the context of IoT security, where labeled attack data is often scarce, hindering the effective development of ML-IDS. Although state-of-the-art ML-IDS solutions achieve high accuracy in detecting prevalent attack types, they often exhibit low precision when faced with rare or infrequent attacks \cite{macas2022survey}. This limitation is particularly concerning in the context of zero-day attacks, which, by definition, are absent in training data, thereby posing significant challenges for effective detection. Figure~\ref{fig:unknown_attacks} illustrates the limitations of current supervised ML-IDS methods in handling zero-day attacks, revealing a substantial decline in performance when faced with previously unseen threats. Self-Supervised Learning (SSL) presents a promising solution by leveraging only benign data to learn generalized representations of normal behavior, enabling the model to detect anomalies as deviations from this learned baseline \cite{nakip2024online}. Unlike supervised learning, SSL does not require manually labeled attack data. 
This property is particularly beneficial in IoT environments, where attack data is scarce, and manual labeling is both time-consuming and costly. By training exclusively on benign data, SSL-based approaches enhance resilience to shifts in attack distributions and exhibit improved generalization to previously unseen threats. However, existing SSL-IDS solutions \cite{caville2022anomal, yue2022contrastive, nguyen2023ts} fail to account for the continuously evolving nature of both malicious and benign behaviors. As a result, these approaches frequently experience a decline in detection performance over time due to catastrophic forgetting, wherein previously acquired knowledge is lost during model updates.

\begin{figure}
    \centering
    \includegraphics[width=\linewidth]{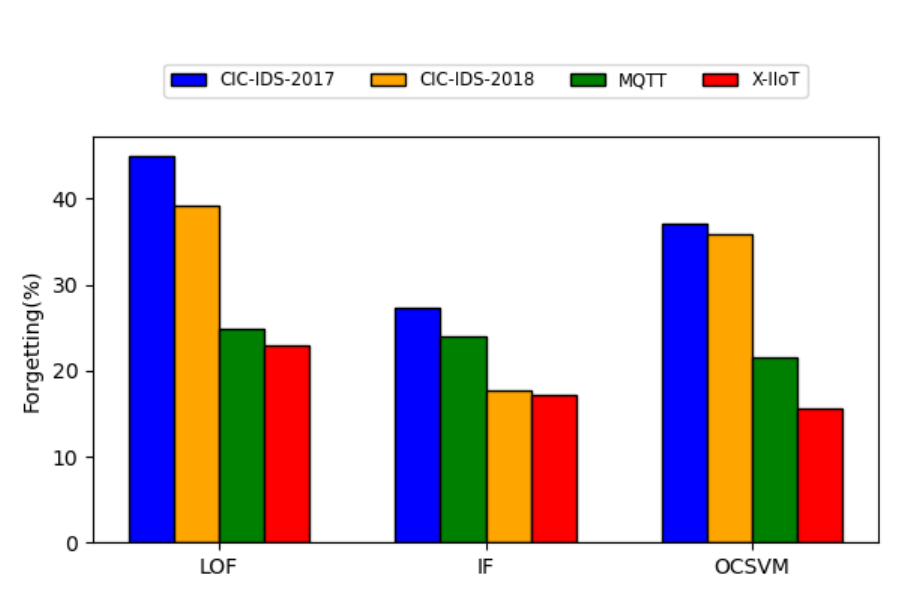}
    \caption{Forgetting performance of state-of-the-art models}
    \label{fig:forgetting_motivation}
\end{figure}

Beyond the scarcity of labeled attack data, a second critical challenge in IoT networks lies in their inherently dynamic and rapidly evolving nature. Both legitimate usage patterns and adversarial behaviors evolve over time, giving rise to a phenomenon known as concept drift \cite{wahab2022intrusion}. Most existing ML-IDS solutions adopt a static, train-and-deploy paradigm, which limits their ability to detect newly emerging threats or adapt to evolving patterns of benign behavior \cite{verwimp2023continual}. Moreover, when these models are updated with new data, they frequently experience \textit{catastrophic forgetting}, a phenomenon in which the integration of new knowledge leads to the unintentional loss of previously learned information \cite{de2021continual}. As shown in Figure~\ref{fig:forgetting_motivation}, state-of-the-art anomaly detection methods—Local Outlier Factor (LOF), Isolation Forest (IF), and One-Class Support Vector Machine (OCSVM)—exhibit significant performance degradation on earlier tasks as new tasks are introduced, underscoring the severity of forgetting in current ML-based intrusion detection systems. In this figure, larger values on the vertical axis correspond to increased forgetting, underscoring the need for models capable of retaining prior knowledge across evolving tasks. 

Continual learning (CL) has emerged as a promising paradigm for overcoming this limitation, enabling models to incrementally learn from evolving data streams while preserving previously acquired knowledge~\cite{wang2024comprehensive}. However, most CL-IDS solutions \cite{amalapuram2021handling, faber2024lifelong, karn2021learning, zhang2024continual, prasath2022analysis, channappayya2023augmented} remain grounded in supervised learning, necessitating the availability of labeled data, which is often impractical in real-world IoT environments. Integrating SSL into CL offers a promising solution, allowing for continuous adaptation without the need for explicit attack labels. The model continuously learns from evolving benign data streams through SSL objectives, updating its representations without sacrificing previously acquired knowledge.

To tackle both the scarcity of labeled data and the non-stationarity of IoT traffic, we propose CITADEL (Continual Anomaly Detection for Enhanced Learning in IoT Intrusion Detection), a unified framework that integrates self-supervised and continual learning for effective anomaly detection (Figure~\ref{fig:citadel-framework}). CITADEL consists of three key modules: (1) a tabular-to-image transformation module that prepares network data for visual modeling; (2) a memory-aware representation learner that employs continual learning strategies to mitigate catastrophic forgetting, using masked autoencoders trained exclusively on benign data; and (3) a novelty detector that distinguishes between benign and anomalous inputs. 

In this work, we demonstrate that our model not only achieves competitive average detection performance relative to state-of-the-art methods but also substantially mitigates the degradation of performance on previously learned tasks as new data is incorporated. The retention of prior knowledge is particularly crucial in intrusion detection scenarios, where previously encountered attack types are likely to reoccur during real-world IoT operation~\cite{fuhrman2025cnd}. Furthermore, our approach facilitates accelerated adaptation to emerging threat types, thereby enhancing the detection of zero-day attacks through improved knowledge transfer from previously learned tasks. Experimental results on real-world IoT intrusion datasets demonstrate that CITADEL outperforms state-of-the-art self-supervised continual learning methods, achieving up to a 72.9\% improvement over the task-agnostic VAE-based lifelong anomaly detector (VLAD) \cite{faber2023vlad} in mitigating forgetting, while maintaining consistent performance across evolving scenarios.
 \section{Related Work}\label{sec:background}

\subsection{IoT Security and Intrusion Detection}
The increasing integration of Internet of Things (IoT) devices into daily life has introduced significant benefits in terms of convenience and efficiency, but it has also given rise to substantial security challenges. These devices frequently operate in resource-constrained environments, utilize a variety of communication protocols, and produce heterogeneous traffic patterns, all of which complicate the application of traditional security measures \cite{alwaisi2024securing}. In response to the rising frequency and increasing sophistication of cyber attacks targeting IoT systems, Intrusion Detection Systems (IDS) have emerged as a critical defensive solution, monitoring network traffic and device behavior to detect and mitigate security threats \cite{hassan2023intrusion}. However, conventional signature-based IDS often struggle to cope with the unique characteristics and dynamic behaviors inherent in IoT networks \cite{khan2022deep}. 
This has led to growing interest in machine learning-based IDS (ML-IDS), which are particularly well-suited for IoT environments due to their ability to automatically learn patterns from data, reduce the reliance on manually crafted rules, and adapt to emerging threats \cite{chen2025dynamite}. 

Among these approaches, supervised learning methods such as Random Forest and Neural Networks have demonstrated strong performance in detecting known attack patterns, provided that ample labeled data are available for training \cite{malele2023testing}. However, their effectiveness is substantially limited in the presence of zero-day attacks or infrequent threat behaviors, as they depend heavily on prior knowledge encoded in labeled datasets \cite{caville2022anomal}. On the other hand, unsupervised learning methods such as Isolation Forest and k-means clustering operate without labeled data, enabling the detection of previously unseen attacks. 
However, these models often exhibit high false positive rates and limited generalization in complex, high-dimensional network environments, and they tend to overfit to training patterns, reducing their robustness to distribution shifts \cite{verkerken2022towards, almaraz2023enhancing}. The dynamic and evolving nature of IoT traffic further complicates these challenges, underscoring the need for adaptive and scalable intrusion detection approaches.

\subsection{Self-supervised Learning for Intrusion Detection}
Self-Supervised Learning (SSL) has emerged as a promising solution that bridges the gap between supervised and unsupervised learning by leveraging unlabeled data to learn useful representations, reducing reliance on manual labeling \cite{li2025safe}. 
SSL generates supervisory signals directly from the data by employing carefully designed pretext tasks, such as masked input prediction, transformation recognition, or contrastive learning based on different views of the same sample \cite{balestriero2023cookbook}. These tasks enable the model to learn rich and generalizable feature representations without requiring supervision. In the context of intrusion detection, SSL methods are typically trained solely on benign data to learn a baseline of normal behavior, enabling the identification of deviations as potential anomalies. As a result, SSL can effectively leverage the abundance of benign traffic in IoT networks, providing greater scalability and adaptability to novel attack patterns compared to traditional machine learning algorithms \cite{kour2025self}. 

Recent literature has introduced a variety of approaches for applying SSL to intrusion detection \cite{almaraz2023enhancing, yang2023malicious, caville2022anomal, nguyen2023ts}. Some methods apply contrastive learning to model network behavior, such as transforming traffic data into images for pre-training with contrastive objectives \cite{almaraz2023enhancing}, or designing masking strategies and architectural enhancements to improve representations \cite{yang2023malicious, yue2022contrastive, wang2023robust}. Others leverage graph neural networks to capture structural properties of IoT traffic and apply various SSL objectives to model entity-level or communication-level behaviors \cite{nguyen2023ts, caville2022anomal}. A third line of work focuses on representation learning for anomaly detection, employing SSL to extract robust features for downstream novelty detectors like Isolation Forest or their deep variants \cite{alghushairy2020review, al2021isolation, xu2023deep}. Despite this progress, these SSL methods rely on fixed pretraining phases and lack mechanisms for continual adaptation. As a result, they struggle to maintain performance in non-stationary environments, where both benign and malicious behaviors evolve over time. In contrast, our approach incorporates self-supervised learning within a continual learning framework, enabling incremental adaptation to evolving benign traffic without requiring any labeled attack data.

\subsection{Continual Learning for Intrusion Detection}

Most ML-IDS solutions operate under the assumption that data is independent and identically distributed, which fails in real-world IoT settings where both benign and malicious behaviors evolve continuously \cite{fuhrman2025cnd}. This non-stationarity leads to performance degradation, as static models struggle to adapt to novel patterns. Although retraining and online learning provide partial solutions by enabling model updates with new data, they are computationally inefficient and susceptible to \textit{catastrophic forgetting}, where the incorporation of new information can overwrite previously learned knowledge \cite{wu2024mitigating}. To address these challenges, continual learning (CL) has emerged as a promising approach that enables models to learn from streaming data while preserving prior knowledge~\cite{zhang2024continual}. 

CL has recently been explored in the context of intrusion detection. Karn et al. \cite{karn2021learning} introduce a progressive learning framework for cyber threat detection, providing a structured analysis of catastrophic forgetting and evaluating neural network capacity for continual task integration. Oikonomou et al. \cite{oikonomou2023multi} propose a multi-class continual learning-based IDS using a CNN backbone, capable of simultaneously detecting and classifying diverse network threats with high recall and low false alarm rates. Kumar et al. \cite{kumar2023augmented} introduce an enhanced memory-based continual learning method incorporating a novel perturbation-assisted parameter approximation approach, aimed at improving the scalability and overall performance of continual learning models. However, these solutions rely on labeled attack data, which is often unavailable in real-world IoT environments where threats evolve continuously.

Few studies systematically explored the integration of SSL and CL. Ashfahani et al. \cite{ashfahani2022unsupervised} introduced the Autonomous Deep Clustering Network (ADCN), a self-supervised architecture that incrementally adapts clustering structures without requiring full supervision. Within the IDS domain, Zhang et al. \cite{zhang2024continual} proposed a new method which addresses concept drift by dynamically refreshing the memory buffer through strategic sample selection and strategic forgetting. Faber et al. \cite{faber2023vlad} proposed a VAE-based framework that compresses historical features into global prototypes, enabling memory-efficient continual anomaly detection. However, these approaches often lack task structuring or depend on heuristics, limiting their adaptability and representation retention.

In contrast to prior CL-IDS methods, our approach introduces a dedicated self-supervised backbone tailored to evolving intrusion patterns. Rather than relying on standard replay mechanisms or memory heuristics, we propose a unified framework that integrates task-driven concept formation through clustering, metric-based matching of normal and anomalous data, and optimized memory management. Our use of a masked autoencoder (MAE) facilitates feature learning that complements strategic sampling, controlled forgetting, and a hierarchical memory structure. This design enables task-aware adaptation and robust knowledge retention, supporting novel threat detection and offering a scalable solution for dynamic, label-scarce IoT environments.

\begin{figure*}[t]
    \centering
    \includegraphics[width=.95\linewidth]{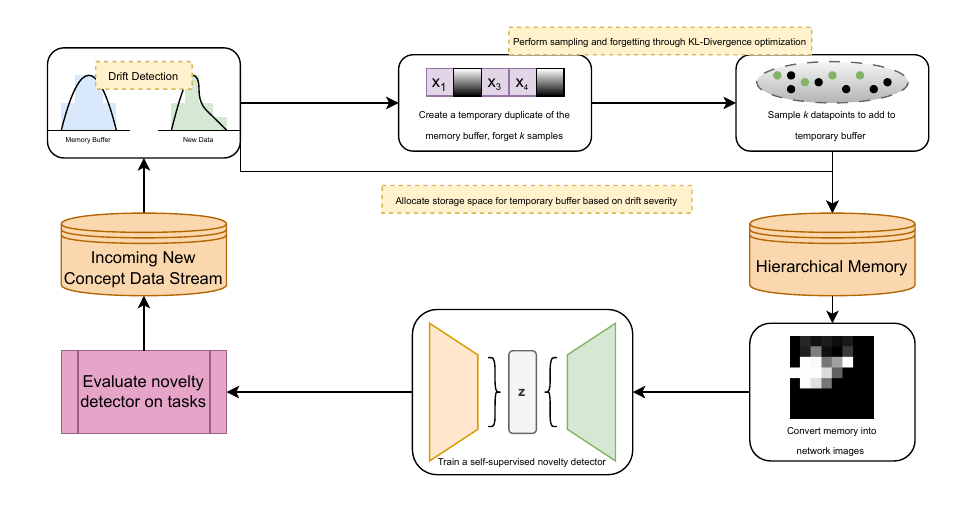}
    \caption{Proposed Anomaly Detection Framework (CITADEL). Upon detecting drift in incoming data, a temporary memory buffer is created for optimized forgetting and strategic sampling. This temporary buffer is then integrated into a hierarchical memory structure, which serves as the training input for the self-supervised novelty detection model.}
    \label{fig:citadel-framework}
\end{figure*}

\section{Proposed Framework: CITADEL}\label{sec:proposal}
A high-level overview of our framework (CITADEL) is illustrated in Figure \ref{fig:citadel-framework}, emphasizing the integration of self-supervised learning and continual learning components. As new data arrives, CITADEL continuously updates a memory buffer containing both recent and past samples. This update process is formulated as an optimization problem that minimizes the Kullback-Leibler divergence between the distribution of incoming data and that of the existing memory buffer. Based on this optimization, new samples are incorporated into a hierarchical memory structure that organizes data according to their dissimilarity, ensuring a representative and diverse training set. The memory buffer serves as input to our self-supervised learning pipeline, which begins by transforming tabular network intrusion data into image representations. These images are then used to train a masked autoencoder (MAE) that learns a latent space representation of the data. Ultimately, a novelty detection head operates on this latent space to identify potential threats. In the following subsections, we provide a detailed description of the self-supervised learning and continual learning components of our framework.

\subsection{Self-supervised Learning}
\begin{figure}[]
    \centering
    \includegraphics[width=1\linewidth]{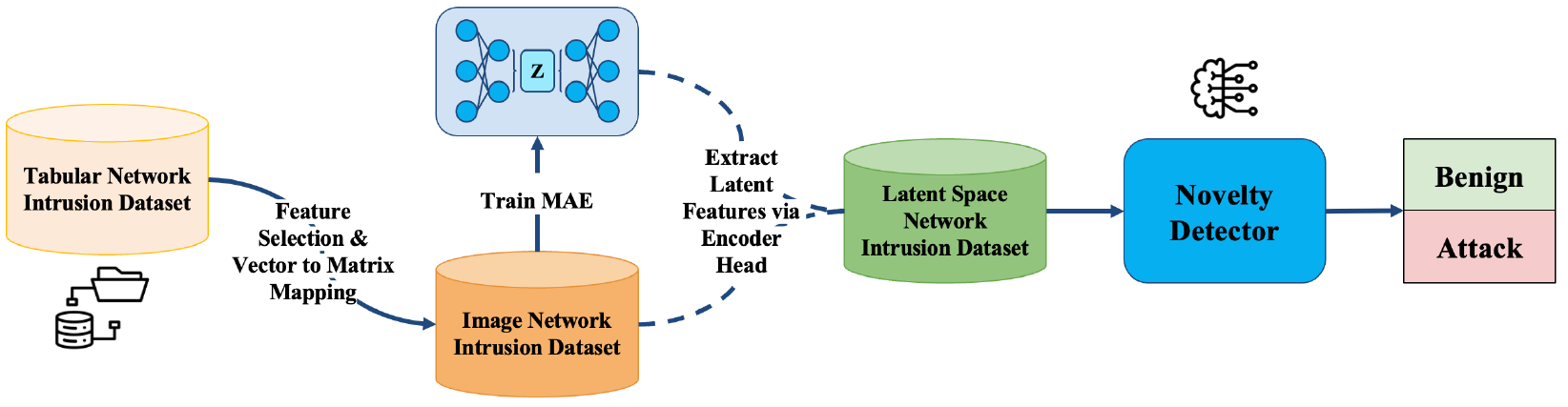}
    \caption{Self-supervised Learning Component (SAFE)}
    \label{fig:ssl-framework}
\end{figure}

Figure \ref{fig:ssl-framework} presents the self-supervised component (SAFE) of our framework, which is composed of four main modules: feature selection, vector-to-image matrix feature mapping, masked autoencoder (MAE) training, and MAE feature extraction-based novelty detection. Motivated by the success of MAEs within the vision domain from recent years \cite{han2024efficient}, our core idea is to transform the original vector features into a suitable matrix, where a geometric relationship exists between each element. This is analogous to the properties of images. Since network intrusion data is inherently tabular, a method is needed to transform it into an image-like format to leverage the MAE vision capabilities. After this transformation, we train an MAE and extract features from its encoder. These features are then used to train a novelty detector, which distinguishes between attack and normal samples. Given a new sample from the test data, we use the trained MAE and novelty detector during inference to determine whether it is normal or an attack. Table \ref{tab:variable-references} introduces variable notations used in this subsection. 

\begin{table}[t]
    \centering
    \small
    \caption{Self-supervised Learning Variable Notations}
    \label{tab:variable-references}
    \scalebox{0.9} {
    \begin{tabular}{ll}
        \toprule
        \textbf{Variable} & \textbf{Description} \\
        \midrule
        \textbf{$\mathcal{D} = \{(x_i,y_i) \: | \: x_i \in \mathbb{R}^d\}_{i=1}^n$}
        & Original Intrusion Dataset \\
        \midrule
        \textbf{$\mathcal{D}_F = \{(x_{fi},y_i) \: | \: x_{fi} \in \mathbb{R}^k, k \leq d\}_{i=1}^n$}
        & Feature Filtered Dataset \\
        \midrule
        \textbf{$\mathcal{D}_I = \{(x_i',y_i) \: | \: x_i' \in \mathbb{R}^{d' \times d'}\}_{i=1}^n$} 
        & Image Intrusion Dataset  \\
        \midrule
        \textbf{$\mathcal{D}_L = \{(x_i'',y_i) \: | \: x_i'' \in \mathbb{R}^{d''}\}_{i=1}^n$} 
        & Latent Space Dataset  \\
        \midrule
        \textbf{$0$} 
        & Label for Normal Data  \\
        \midrule
        \textbf{$1$} 
        & Label for Attack Data  \\
        \bottomrule
    \end{tabular}}
\end{table}

\subsubsection{Feature Selection} Selecting an effective feature selection algorithm is crucial for reducing dimensionality, thereby facilitating the pre-training of an MAE with reduced computational overhead. Moreover, given that the latent space generated by the MAE inherently possesses fewer dimensions than the original feature space, it is imperative to ensure that only the most salient features contribute to its construction. We utilize a key property of principal component analysis (PCA) to perform dimensionality reduction \cite{song2010feature}. In addition to its low computational cost, we specifically choose PCA because it is an unsupervised technique, eliminating the need for prior knowledge of network intrusions. Consider $\mathcal{D}_{train} = \{(x_{i_{train}},y_{i_{train}}) \: | \: x_i \in \mathbb{R}^{d_{train}}, y_{i_{train}} = 0\}_{i=1}^{n_{train}}$ to be the training set of our original dataset. 
Since every principal component $PC_i$ is a linear combination of the original features $x_{i_{train}}$, denote it by $PC_j = a_{j,1}x_{1} + a_{j,2}x_{2} + a_{j,3}x_{3} + ... + a_{j,d_{train}}x_{j,d_{train}}$ where $j$ indicates the $j$th principal component, and $a_{j,i}, i \in \{1,...,d_{train}\}$ are the loadings of each feature for the respective $j$th principal component. We then calculate the ranking of each $x_{i_{train}}$ through $\mathcal{R}(x_{i_{train}}) = \sum_{j=1}^M |a_{j,i}|$, where $M$ is the total number of principal components. To select $M$, we pick the least number of principal components which cumulatively account for an explained variance ratio, the latter of which is set to $95\%$. Applying $\mathcal{R}(x_{i_{train}})$ to each feature, we obtain a numerical set that can be sorted in descending order to represent the importance of each feature. Select an arbitrary $k \in \mathbb{N}, 1 \leq k \leq d$, and retain only the top $k$ features ranked by $\mathcal{R}(x_{i_{train}})$ to obtain $\mathcal{D}_F$.

\subsubsection{Vector to Image Matrix Mapping} As discussed earlier, MAE benefits have been demonstrated in the vision domain; therefore, it is necessary to train the model on an image dataset. Since network intrusion data is tabular, we need to create a mapping for each vector datapoint into a subsequent image matrix counterpart. This means that each feature within the transformed image matrix should have a spatial relationship with neighboring features, which outlines our goal to transform from $\mathcal{D}_F$ to $\mathcal{D}_I$. \par 
We leverage DeepInsight \cite{9701572} to efficiently transform the feature vectors into feature matrices and create a map between each datapoint $(x_i, y_i)$ and its respective transformation $(x_i',y_i)$. 
To achieve this, our configuration of DeepInsight utilizes t-Stochastic Neighbor Embedding (t-SNE) \cite{van2008visualizing} to reduce each feature's vector representation from $\mathbb{R}^n$ to $\mathbb{R}^2$. This allows us to find the $\mathbb{R}^2$ coordinate locations of each specific feature, $x_j$, which we can denote as $(a_j, b_j)$. If a bijection between features and their coordinates does not exist, discretization techniques such as linear sum assignment are applied to establish a one-to-one correspondence, ensuring that the coordinates for our features remain unique. DeepInsight then applies a convex hull algorithm to generate the largest polygon that encapsulates all of the $\mathbb{R}^2$ feature representation points. Performing this, along with then reshaping the data into a $d' \times d'$ square matrix, creates a finite space that contains every feature representation whilst having the smallest area (thereby reducing the number of irrelevant, empty coordinates). Then, we iterate through our datapoints and for each $(x_{fi},y_i)$ map every feature $x_{fi,s} \in x_{fi}, s \in \{1,...,k\}$ to $(a_j,b_j)$ of the $\mathbb{R}^{d' \times d'}$ matrix in accordance to the original intensity value, normalized as an integer between $[0,255]$ (the standard numerical color value range). After applying the t-SNE based tabular image conversion (trained on $\mathcal{D}_{F_{train}} = \{(x_{fi_{train}},y_{i_{train}}) \: | \: x_i \in \mathbb{R}^{k_{train}}, y_{i_{train}} = 0\}_{i=1}^{n_{train}}$) to the entire dataset of $\mathcal{D}_F$, we obtain $\mathcal{D}_I$ (image intrusion dataset).
\par 

\subsubsection{MAE Training} Given that $\mathcal{D}_I$ consists of matrices with spatial relationships between entries, it is well-suited for use as a vision dataset. Our aim is to pre-train an MAE with $\mathcal{D}_{I_{train}}=\{(x'_{i_{train}},y_{i_{train}}) \: | \: x'_{i_{train}} \in \mathbb{R}^{d' \times d'}\}_{i=1}^{n_{train}}$ to learn its latent representation. 
For our MAE architecture, we designed an encoder-decoder structure that processes inputs through convolutional layers and randomly masks training inputs according to a fixed ratio. Each training datapoint was reconstructed after the mask was applied, and the loss for the reconstructed image was then evaluated through the mean squared error (MSE) of individual masked features between the original datapoint and the reconstructed one. 
We also construct a bottleneck layer, known colloquially as the latent space, which serves as the final layer of the encoder and the input layer to the decoder.  
Then, we train our MAE on $\mathcal{D}_{I_{train}}$. Consistent with the typical behavior of autoencoders, it learns a compact and latent representation of the features that is provided to our novelty detector. 

\subsubsection{Feature Extraction and Novelty Detection} 
Given the trained MAE, let us consider its encoder head $\mathcal{H}$. In this module, we explain how we use $\mathcal{H}$ to obtain $\mathcal{D}_L$ (latent space features), and then subsequently apply a novelty detector as our penultimate step to classify attacks. Since $\mathcal{H}$ outputs to a latent space, we effectively input $\mathcal{D}_I$ through the encoder, which will take every image datapoint inside and convert them into a latent vector. Combining these individual latent vector datapoints, they collectively form $\mathcal{D}_L$. We can use these new features to train a novelty detector. Specifically, consider $\mathcal{D}_{L_{train}} = \{(x''_{i_{train}},y_{i_{train}}) \: | \: x''_{i_{train}} \in \mathbb{R}^{d''}, y_{i_{train}} = 0\}_{i=1}^{n_{train}}$. We use $\mathcal{D}_{L_{train}}$ to train a novelty detector, $\mathcal{C}$, which serves as our final attack classifier. We specifically chose $\mathcal{C}$ to be Local Outlier Factor (LOF) \cite{cheng2019outlier}. This decision was based on the fact that LOF identifies anomalies by assessing local density deviations, rather than relying on global distance metrics, thereby facilitating anomaly detection through the evaluation of a data point’s density relative to its local neighborhood. In the latent space where only essential feature representations are preserved, the local structure surrounding each datapoint becomes clearer and more compact, thereby improving the efficacy of LOF. Additionally, LOF exhibits a computationally efficient runtime compared to many other anomaly detection methods, rendering it particularly well-suited for IoT intrusion detection. \par

To address the challenges of non-stationary data, evolving attack patterns, and knowledge retention in dynamic IoT environments, we extend our SAFE framework with a novel continual learning solution for adaptive intrusion detection.

\begin{table}[]
    \centering
    \small
    \caption{Continual Learning Variable Notations}
    \label{tab:variable-references-cl}
    \scalebox{0.9} {
    \begin{tabular}{ll}
        \toprule
        \textbf{Variable} & \textbf{Description} \\
        \midrule
        \textbf{$X_{\text{new}} \in \mathbb{R}^{m \times d}$} & New concept data of $m$ samples and $d$ features \\
        \midrule
        \textbf{$\mathcal{M} \in \mathbb{R}^{l \times d}$} & Hierarchical memory buffer with size limit $l$\\
        \midrule
        \textbf{$\mathcal{M}_{temp} \in \mathbb{R}^{l \times d}$} & Temporary memory buffer for forgetting and sampling  \\
        \midrule
        \textbf{$H_i \in \mathbb{R}^{b \times d}$} & Distribution of $i$-th feature values grouped into $b$ bins\\
        \midrule
        \textbf{$\mathbf{w}_s \in [0,1]^l$} & Weight vector to optimize for strategic sampling\\
        \midrule
        \textbf{$\mathbf{w}_f \in [0,1]^m$} & Weight vector to optimize for strategic forgetting\\
        \bottomrule
    \end{tabular}}
\end{table}

\subsection{Continual Learning}\label{sec:proposal}
\begin{figure}[]
    \centering
    \includegraphics[width=1\linewidth]{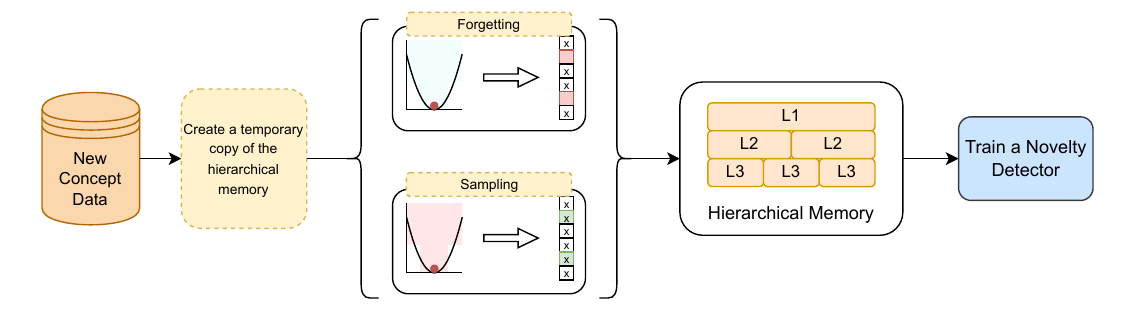}
    \caption{CITADEL Continual Learning Component}
    \label{fig:cl-framework}
\end{figure}

Integrating self-supervised and continual learning enables scalable, label-efficient adaptation to novel IoT threats, but poses challenges such as catastrophic forgetting and representation drift, which our framework addresses. Figure \ref{fig:cl-framework} illustrates CITADEL's continual learning component, which incorporates concept creation, strategic sampling and forgetting mechanisms, and a hierarchical memory architecture. Our CL component is memory-based and designed to retain representative information over time. It maintains a compact set of informative samples through a sample selection strategy, while dynamically adjusting the memory buffer to prioritize novel data that deviates from previously stored representations. This design draws on insights from prior work on sample retention~\cite{zhang2024continual} and memory efficiency~\cite{faber2023vlad}, but introduces a unified mechanism tailored for lifelong anomaly detection. Table \ref{tab:variable-references-cl} presents variable notations used in this subsection.

\subsubsection{Concept Creation} A concept refers to a collection of samples representing new, incoming data streams, designed to simulate real-world scenarios where data distributions are inherently heterogeneous. To represent this in our experiments, the dataset $\mathcal{D}$ is first partitioned into normal samples $\mathcal{D}_\text{norm}$ and anomaly samples $\mathcal{D}_\text{anom}$, where labels are binarized as $y = 0$ for normal and $y = 1$ for anomalies. Both subsets are clustered into $c$ disjoint concepts using k-means clustering \cite{Faber_2024}. Let $\Phi = \{\phi_1, \dots, \phi_c\}$ and $\Gamma = \{\gamma_1, \dots, \gamma_c\}$, where $\phi_i$ contains the normal samples and $\gamma_i$ stores the anomalous data. Next, we construct tasks as pairs \((\phi_i, \gamma_i)\), each representing a data stream characterized by closely related feature distributions. To form concept-aligned tasks, we apply a greedy matching procedure that pairs each normal concept with its closest anomaly concept based on the Euclidean distance between their centroids \cite{Faber_2024}. This ensures that each task contains a metric-related pair of normal and anomaly samples. The pseudocode for this procedure is shown in Algorithm~\ref{alg:concept-matching}. 

\begin{algorithm}[H]
\caption{Concept Matching Algorithm}
\label{alg:concept-matching}
\begin{algorithmic}[1]
\REQUIRE Normal concepts $\Phi = \{\phi_1, \dots, \phi_c\}$, anomaly concepts $\Gamma = \{\gamma_1, \dots, \gamma_c\}$
\ENSURE Matched tasks $\mathcal{T}_1, \dots, \mathcal{T}_c$
\STATE Initialize task list $\mathcal{T} \leftarrow [\ ]$
\STATE Initialize set of unmatched anomaly concepts $\Gamma_{\text{rem}} \leftarrow \Gamma$
\FOR{$i = 1$ to $c$}
    \STATE Compute centroid $\mu_{\phi} \leftarrow \frac{1}{|\phi_i|} \sum_{x \in \phi_i} x$
    \FOR{each $\gamma_j \in \Gamma_{\text{rem}}$}
        \STATE Compute centroid $\mu_{\gamma_j} \leftarrow \frac{1}{|\gamma_j|} \sum_{x \in \gamma_j} x$
        \STATE Compute distance $d_j \leftarrow \|\mu_{\phi} - \mu_{\gamma_j}\|_2$
    \ENDFOR
    \STATE Find closest anomaly: $j^* \leftarrow \arg\min_j d_j$
    \STATE Match: $\mathcal{T}_i \leftarrow (\phi_i, \gamma_{j^*})$
    \STATE Remove $\gamma_{j^*}$ from $\Gamma_{\text{rem}}$
\ENDFOR
\RETURN $\{\mathcal{T}_1, \dots, \mathcal{T}_c\}$
\end{algorithmic}
\end{algorithm}

\subsubsection{Strategic Sampling and Forgetting} 
In memory-based CL methods, the objective is to carefully select data samples that are representative of the unseen data's distribution, limited by a memory size constraint. Without careful selection, arbitrary data sampling can produce a training set that poorly reflects the true distribution. The first component of our CL framework addresses this idea by solving two constrained optimization problems, detailed as follows:

\textbf{a. Forgetting.} This process starts with new concept data $X_{\text{new}}$ and a buffer $\mathcal{M}_{temp}$, which is a temporary duplicate of our memory buffer $\mathcal{M}$ of $l$ samples and $d$ features. The key idea is that we will optimally forget and sample new data into $\mathcal{M}_{temp}$ before adding it to $\mathcal{M}$. Optimized forgetting attempts to solve the mathematical formulation:
\[
\min_{\mathbf{w_f} \in [0.5,1]^l}\sum_{j=1}^d KL(H_j^{\text{new}} || H_j^{\mathcal{M}_{temp}}(\mathbf{w_f}))
\]
Here, $H_j^{\text{new}}$ and $H_j^{\mathcal{M}_{temp}}(\mathbf{w_f})$ denote the binned frequency distributions of the $j$-th feature from the new data and the current memory buffer, respectively, both discretized into $b$ bins. The memory buffer distribution $H_j^{\mathcal{M}_{temp}}(\mathbf{w_f})$ is computed by weighting each sample's contribution to its corresponding bin according to the weight vector $\mathbf{w_f} \in [0,1]^l$. This optimization problem can be solved using iterative algorithms, and the solution yields an optimal vector $\mathbf{w_f}^*$, where each entry $\mathbf{w_f}^*_i$ reflects the relative importance of sample $i$ in preserving the feature-wise distributional similarity between $\mathcal{M}_{temp}$ and $X_{\text{new}}$. Next, we define $\mathcal{I}_{\text{drop}}$ as the indices of all samples in the memory buffer whose optimized weights satisfy $\mathbf{w_f}^*_i < 0.5$. If fewer than $k$ such samples exist, we select additional samples with the lowest remaining weights in $\mathbf{w_f}^*$ to reach the desired forgetting quota size $k$. The memory buffer is then updated as
\[
\mathcal{M}_{temp} \leftarrow \mathcal{M}_{temp} \setminus \{x_i \mid i \in \mathcal{I}_{\text{drop}}\}
\]
\textbf{b. Sampling.} After removing outdated samples, we start the re-population process by using new data drawn from $X_{\text{new}} \in \mathbb{R}^{m \times d}$. This yields the following optimization problem:
\[
\min_{\mathbf{w_s} \in [0,1]^m} \sum_{j=1}^d KL\left(H_j^{\text{new}} \,\bigg\|\, \frac{H_j^{\mathcal{M}_{temp}} + H_j^{\text{new}}(\mathbf{w_s})}{2}\right)
\]
where $H_j^{\text{new}}(\mathbf{w_s})$ is the weighted histogram (with $b$ bins) of the $j$-th feature from $X_{\text{new}}$, constructed by applying the weight $\mathbf{w_s}_i$ to each sample. The mean $\frac{H_j^{\mathcal{M}_{temp}} + H_j^{\text{new}}(\mathbf{w_s})}{2}$ provides a symmetrized estimate of the merged distribution, promoting balance between old and new data. The problem is solved via iterative optimization, yielding an optimal vector $\mathbf{w_s}^*$.
We then select the $k$ samples in $X_{\text{new}}$ with the highest weights:
\[
\mathcal{I}_{\text{select}} = \text{indices of the } k \text{ largest entries in } \mathbf{w_s}^*
\]
and update the temporary buffer:
\[
\mathcal{M}_{temp} \leftarrow \mathcal{M}_{temp} \cup \{x_i \mid i \in \mathcal{I}_{\text{select}}\}
\]
\subsubsection{Hierarchical Memory} We also utilize a hierarchical memory structure, a common CL component to allocate appropriate memory sizes based on how much new data deviates from existing data. This component organizes memory into levels of importance, where lower levels are allocated more space and store higher-priority data. Each concept is stored at an appropriate level based on the severity of distributional drift from the previous memory, and memory across levels is continuously summarized to meet the global memory constraint. While the forgetting and sampling steps focused on finding new representative samples and keeping important older samples, adding $\mathcal{M}_{temp}$ to an appropriate level in $\mathcal{M}$ allows us to further downsize the new incoming data and only keep the important ones. When new concept data arrives, we apply the Kolmogorov-Smirnov test~\cite{berger2014kolmogorov} to check if the distribution of $X_{\text{new}}$ has drifted from the distribution of $\mathcal{M}$ with a significance level of $\alpha=0.05$. If drift is detected, we then calculate a severity score $s$ by computing the average of the Kolmogorov-Smirnov statistics for each feature $i \in \{1,...,d\}$. If $K_i$ is the statistic for feature $i$, then this is
\[
s = \frac{1}{d}\sum_{i=1}^d K_i
\]
We then utilize a function $\text{level}(s) : [0,1] \to \mathcal{L}$ that smoothly maps $s$ to an assigned level via an exponential transformation:
\[
\text{level}(s) = \left\lfloor \mathcal{L}_{\min} + (\mathcal{L}_{\max} - \mathcal{L}_{\min}) \cdot \frac{e^{\lambda s} - 1}{e^{\lambda} - 1} + 0.5 \right\rfloor
\quad
\]
where $\lambda > 0$ is chosen by the user and levels $\mathcal{L} = \{1,...,L\}$. Allocation decays geometrically with level index $\mathcal{L}$, adhering to the following:
\[
\text{Alloc}(\mathcal{L}) = |\mathcal{M}|\cdot \frac{1/\gamma^{\mathcal{L}-1}}{\sum_{j=1}^L 1/\gamma^{j-1}}, \gamma > 1
\]
Once the new data has gone through the optimized forgetting and sampling process, we ensure that it fits within its assigned level's memory allocation by downsizing it. This process is done by applying K-means clustering to generate $c$ cluster centroids, selecting Alloc$(\mathcal{L})$ samples closest to any centroid (through Euclidean distance) to be added into the appropriate level in $\mathcal{M}$. In the case where different data groups are assigned to the same level, we can resolve it by uniformly allocating the same amount of memory to each updated buffer for that specific level. Furthermore, to ensure that $\mathcal{M}$ always contains $l$ samples, we use the first concept to populate the first level, and then any levels afterward whilst marking them as they do not inherently belong there. When we populate $\mathcal{M}$ as the iterations pass, they will be assigned to a level within $\mathcal{M}$. If that level is not the first level, we will phase out the marked data and replace it with the new samples.
\section{Experimental Analysis}\label{sec:results}
\subsection{Experimental Setup}
\subsubsection{Hardware Setup} Experiments ran on a Linux server with 32 CPUs, 64 GB RAM, and an NVIDIA A100 GPU. 

\subsubsection{Self-Supervised Learning Setup} 
\begin{figure}[t]
    \centering
    \captionsetup{justification=centering}
    \includegraphics[width=1\linewidth]{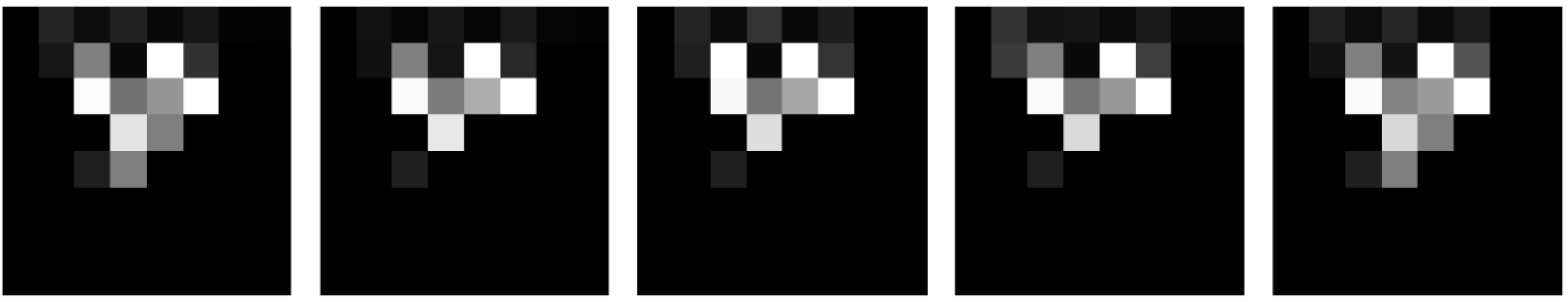}
    \includegraphics[width=1\linewidth]{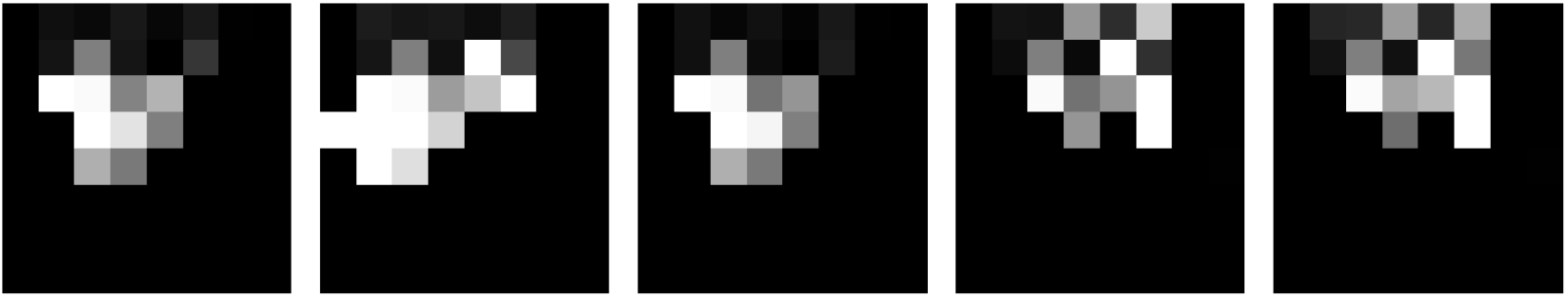}
    \caption{Selected image matrices of normal (top) and attack (bottom) data from X-IIoTID. Normal data image matrices exhibit visual homogeneity, whereas attack data appear in diverse forms, producing distinctly different transformations.}
    \label{fig:xiiot-images}
\end{figure}

We select the top $k=31$ features through PCA feature selection. Then, the selected features are converted into their grayscale image representation. Figure~\ref{fig:xiiot-images} provides example image representations. We chose to set our feature matrix dimension as $8 \times 8$ across all datasets. The MAE is then trained with $20$ epochs, the Adam optimizer, and an MSE loss function for the masked portions. For each datapoint, we randomly mask $75\%$ of its entries.

\subsubsection{Concept Creation Setup}
We standardized each dataset by creating 5 separate concepts. As defined by our framework, each matched pair $\mathcal{T}_i = (\phi_i, \gamma_i)$ in ${\mathcal{T}_1, \dots, \mathcal{T}_5}$ is partitioned into training and testing subsets in a 70/30 ratio:
\[
\phi_i = \phi_i^\text{train} \cup \phi_i^\text{test}, \quad \gamma_i = \gamma_i^\text{train} \cup \gamma_i^\text{test}.
\]
The training and testing sets for task $\mathcal{T}_i$ are defined as:
\[
\mathcal{T}_i^\text{train} = \phi_i^\text{train} \cup \gamma_i^\text{train}, \quad
\mathcal{T}_i^\text{test} = \phi_i^\text{test} \cup \gamma_i^\text{test},
\]
with corresponding labels $y = 0$ for all $x \in \phi_i$ and $y = 1$ for all $x \in \gamma_i$.
At time step $t$, only $\mathcal{T}_t^\text{train}$ is used for model update and evaluation is performed on $\mathcal{T}_j^\text{test}$ for all $j \leq t$, populating a result matrix $R \in \mathbb{R}^{c \times c}$ such that:
\[
R_{i,j} = \text{performance}(\text{model after task } i, \ \mathcal{T}_j^\text{test})
\]
where performance is defined later as a task-specific metric.
\subsubsection{Continual Learning Strategy Setup} All parameter values were selected through rigorous analysis to ensure balanced efficiency and performance across datasets. We initialize the replay buffer $\mathcal{M}$ with a capacity of $l=5000$ samples. As new data $X_{\text{new}}$ arrives, it is processed sequentially as follows:
\paragraph{Strategic Forgetting} Across all datasets, a forgetting quota of size $k=1000$ was set. Upon minimizing the KL-divergence between the histogram of the new data and the weighted histogram of the memory buffer, $1000$ samples corresponding to the lowest weights were removed from $\mathcal{M}$. 
\paragraph{Strategic Sampling} We employ a lower threshold of $k=1000$ samples to add into $\mathcal{M}$. After the optimization problem is solved, $1000$ samples from $X_{new}$ corresponding to the largest weight values become included in our memory. 
\paragraph{Hierarchical Memory} We use $L_{max}=10$ levels to store incoming data concepts and chose $\lambda=5.5$ as the parameter for our level assignment function, allowing granular control of how $X_{new}$, after going through strategic sampling and forgetting, is assigned memory size. 
\subsection{Datasets} We evaluate our approach on six large-scale datasets—including IoT datasets (MQTTset, WUSTL-IIoT, X-IIoTID) and general intrusion detection benchmarks (UNSW-NB15, CIC-IDS 2017, CIC-IDS 2018)—to demonstrate its effectiveness across IoT and broader network environments. Table \ref{datasets} summarizes the key characteristics of these datasets, including the number of features, attack types, and samples. \\
\begin{table}[]
\centering
\caption{Selected intrusion datasets}
\scalebox{0.85}{
\begin{tabular}{|c|c|c|c|c|}
\hline
\textbf{Dataset} & \textbf{Year} & \textbf{\begin{tabular}[c]{@{}c@{}}Number of \\ Features\end{tabular}} & \textbf{\begin{tabular}[c]{@{}c@{}}Number of \\ Attacks\end{tabular}} & \textbf{\begin{tabular}[c]{@{}c@{}}Number of \\ Samples\end{tabular}} \\ \hline
MQTTset~\cite{vaccari2020mqttset}          & 2020          & 33                                                                     & 5                                                                     & 20M                                                                   \\ \hline
WUSTL-IIoT\cite{zolanvari2021wustl}        & 2021          & 41                                                                     & 4                                                                     & 1M                                                                    \\ \hline
X-IIoTID~\cite{al2021x}         & 2021          & 59                                                                     & 18                                                                    & 0.9M                                                                  \\ \hline
UNSW-NB15~\cite{moustafa2015unsw}    & 2015          & 49                                                                     & 9                                                                    & 2.5M     
\\ \hline
CIC-IDS 2017~\cite{Sharafaldin2018TowardGA}    & 2017          & 78                                                                      & 15                                                                    & 8.4M  
\\ \hline
CIC-IDS 2018~\cite{Sharafaldin2018TowardGA}    & 2018          & 72                                                                     & 15                                                                    & 7.8M  
\\ \hline
\end{tabular}}
\label{datasets}
\end{table}
\subsection{Baselines} For our state-of-the-art baseline comparisons, we classify the methods into two primary categories: (1) anomaly detection techniques and (2) end-to-end CL solutions: \\
\subsubsection{Anomaly detection models} We select the following anomaly detection methods as representative baselines: \\
\textbf{LOF~\cite{cheng2019outlier}:} Local Outlier Factor (LOF) calculates the local density deviation of a sample relative to its neighbors, identifying points with significantly lower density than their surroundings as anomalies. \\
\textbf{IF~\cite{liu2008isolation}:} Isolation Forest (IF) isolates anomalies by constructing random decision trees. Anomalies, being easier to isolate than normal points, typically exhibit shorter path lengths within these trees. \\
\textbf{SGD-OCSVM~\cite{nguyen2019scalable}:} Stochastic Gradient One-Class Support Vector Machine (SGD-OCSVM) is a scalable variant of the traditional OCSVM, optimized using stochastic gradient descent. It learns a decision boundary around normal data to distinguish anomalies in high-dimensional settings.\\
\textbf{SLAD~\cite{xu2023fascinating}: }Self-Supervised Learning Anomaly Detection (SLAD) leverages SSL to build a feature representation space, then uses these representations to detect anomalies. \\
\textbf{ICL~\cite{shenkar2022anomaly}: }Instance-conditioned learning (ICL) captures instance-conditioned representations to improve anomaly detection by minimizing class overlap. \\
\textbf{RCA~\cite{liu2021rca}: }Representation-Constrained Autoencoder (RCA) integrates constraints into the autoencoder structure, aiming to learn effective data representations. \\
\textbf{RDP~\cite{wang2019unsupervised}: }Representation learning for Dynamic Predictions (RDP) is designed to learn representations that capture temporal or structured changes within data. \\
\textbf{Anomal-E\cite{caville2022anomal}: }Anomal-E is a framework designed to detect anomalies using graph neural networks (GNNs). It is the first network intrusion detection method to employ GNNs.

\textbf{CL Adaptation with Replay Buffer:} These methods, originally designed for static settings, lack mechanisms for handling non-stationary data. To enable continual learning, we integrate a fixed-size replay buffer $\mathcal{M}$ of length $l$.
\subsubsection{End-to-end CL solutions} These baselines are state-of-the-art continual learning methods proposed in prior works: \\
\textbf{ADCN~\cite{ashfahani2022unsupervised}: }
Autonomous Deep Clustering Network (ADCN) is an unsupervised continual learning model that adapts its network structure over time based on reconstruction bias-variance. It performs clustering in each layer’s latent space and combines these for final predictions. The architecture evolves by adding neurons or layers as needed. \\
\textbf{VLAD~\cite{faber2023vlad}: }VAE-based Lifelong Anomaly Detection (VLAD) is a task-agnostic lifelong anomaly detection framework built upon a variational autoencoder. It detects concept drift using change point detection and stores concepts in a hierarchical memory. Summarized memory supports continual updates via experience replay. \\
\textbf{SSF~\cite{zhang2024continual}: }Strategic Sampling and Forgetting (SSF) detects distributional drift and updates its memory. It forgets outdated samples by minimizing KL divergence between feature distributions. New samples are selected via weighted optimization to best represent the drift. \\
\subsection{Evaluation Metrics}
For self-supervised learning performance, we report precision, recall, and F1-score as the primary metrics for evaluation. 
To evaluate continual learning performance, we compute three metrics: Lifelong PR-AUC, Backward Transfer (BWT), and Forward Transfer (FWT). These are derived from the results matrix $R \in \mathbb{R}^{c \times c}$, where $R_{i,j}$ denotes the PR-AUC score of the model on task $\mathcal{T}_j$ after training on task $\mathcal{T}_i$.
\[
\text{PR-AUC} = \int_0^1 \operatorname{Precision}(\operatorname{Recall}) \, d\, \operatorname{Recall}
\]
\paragraph{Lifelong PR-AUC} This metric captures the model's overall ability to retain performance over time and averages scores from all test tasks up to each time step. LL PR-AUC serves as a robust metric for anomaly detection in imbalanced settings, a common trait of intrusion detection datasets:
\[
\text{LL PR-AUC} = \frac{1}{\frac{c(c+1)}{2}} \sum_{i=1}^{c} \sum_{j=1}^{i} R_{i,j}.
\]
\paragraph{Backward Transfer (BWT)} BWT quantifies the degree to which learning new tasks interferes with or reinforces performance on previously learned tasks. Positive BWT values imply that acquiring new knowledge benefits past performance, while negative values indicate forgetting:
\[
\text{BWT} = \frac{1}{\frac{c(c-1)}{2}} \sum_{i=2}^{c} \sum_{j=1}^{i-1} \left( R_{i,j} - R_{j,j} \right).
\]
\paragraph{Forward Transfer (FWT)} FWT evaluates a model’s ability to leverage previously acquired knowledge to improve learning on unseen tasks. A higher value suggests better generalization across tasks:
\[
\text{FWT} = \frac{1}{\frac{c(c-1)}{2}} \sum_{i=1}^{c-1} \sum_{j=i+1}^{c} R_{i,j}.
\]
Together, these metrics provide insight into a model's ability to retain past knowledge, avoid catastrophic forgetting, and facilitate knowledge transfer throughout the learning process.

\begin{figure}[t]
    \centering
    \captionsetup{justification=centering}
    \includegraphics[width=1\linewidth]{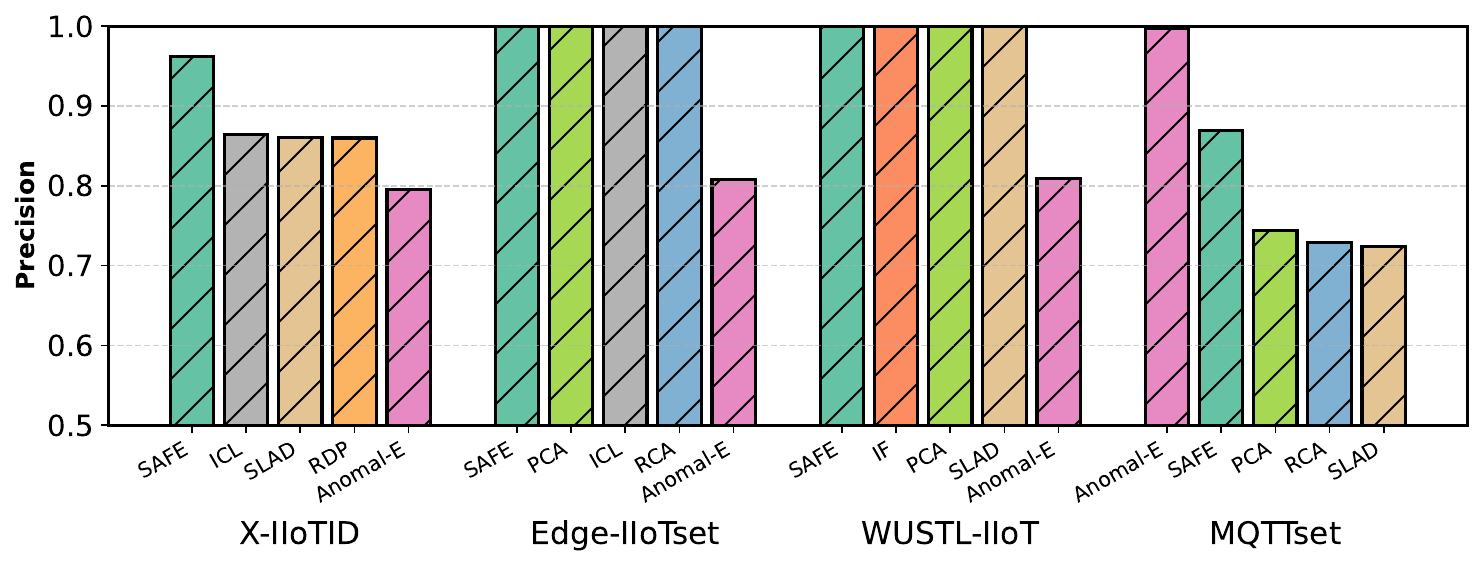}
    \includegraphics[width=1\linewidth]{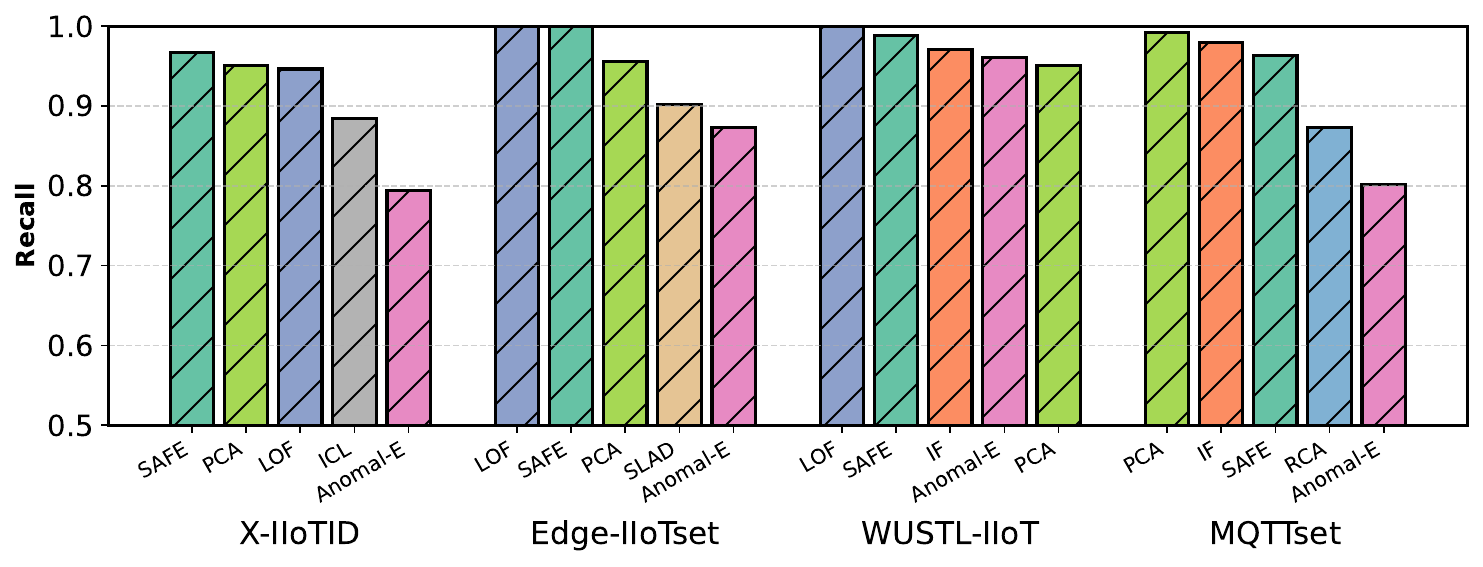}
    \caption{State-of-the-art Precision and Recall Comparison}
    \label{fig:precision-recall}
\end{figure}

\subsection{Results}

\begin{table*}[t]
    \centering
    \captionsetup{justification=centering}
    \caption{SAFE SOTA F1-score Comparison. Bold and underlined cases indicate the best and the second best, respectively.}
    \label{tab:combined-performances}
    \scalebox{0.98}{
    \begin{tabular}{lcccccc}
        \toprule
        \textbf{Model} & \textbf{X-IIoTID} & \textbf{WUSTL-IIoT} & \textbf{Edge-IIoTset} & \textbf{MQTTset} & \textbf{Average} & \textbf{Std. Dev.}\\
        \midrule
        \textbf{SAFE} & \textbf{0.9641} & \textbf{0.9940} & \textbf{0.9995} & \textbf{0.9138} & \textbf{0.9679} & \textbf{0.0392} \\
        \textbf{Anomal-E} & 0.7944 & 0.8654 & 0.8092 & \underline{0.8753} & 0.8361 & \underline{0.0402} \\
        \textbf{LOF} & 0.7414 & 0.9625 & \underline{0.9994} & 0.4469 & 0.7875 & 0.2541 \\
        \textbf{IF} & 0.7209 & \underline{0.9852} & 0.8848 & 0.7979 & 0.8472 & 0.1138 \\
        \textbf{PCA} & \underline{0.8965} & 0.9745 & 0.9773 & 0.8504 & \underline{0.9246} & 0.0620 \\
        \textbf{SLAD} & 0.8571 & 0.9494 & 0.7954 & 0.7244 & 0.8316 & 0.0954 \\
        \textbf{ICL} & 0.8741 & 0.9312 & 0.9473 & 0.5748 & 0.8318 & 0.1742 \\
        \textbf{RCA} & 0.8168 & 0.9474 & 0.9446 & 0.7944 & 0.8758 & 0.0816 \\
        \textbf{RDP} & 0.8628 & 0.9307 & 0.8756 & 0.7148 & 0.8459 & 0.0923 \\
        \bottomrule
    \end{tabular}}
\end{table*}

We first evaluate SAFE, our previously developed model without continual learning, to establish a baseline. We then present the enhanced performance achieved by incorporating continual learning through CITADEL.
\paragraph{SAFE Results} Figure~\ref{fig:precision-recall} presents a comparison of precision and recall performance against state-of-the-art methods. To illustrate their differences, we retain the top four performers from each dataset and include Anomal-E for comparison. In zero-day attack detection, it is often more critical to minimize false negatives, as failing to detect such an attack can lead to severe security breaches. Consequently, a higher false positive rate is generally more acceptable in order to ensure that zero-day attacks are not missed. Remarkably, SAFE consistently ranks within the top three across all datasets in terms of recall, demonstrating its suitability for real-world network intrusion scenarios. Furthermore, SAFE consistently ranks within the top two for precision across all datasets, highlighting its robustness in accurately detecting intrusions. Table \ref{tab:combined-performances} presents a state-of-the-art comparison of F1-scores for each model across all datasets. Notably, a comparison of SAFE results with those of the previously highest-performing method reveals a consistent performance improvement across all datasets: X-IIoTID (+6.76\%), Edge-IIoTset (+0.01\%), WUSTL-IIoT (+0.88\%), and MQTTset (+3.85\%). More importantly, this demonstrates that SAFE outperforms all nine selected models across four datasets, underscoring the effectiveness of SAFE. \\

\begin{table*}[t]
    \centering
    \caption{SOTA LL PR-AUC Comparison. Bold and underlined cases indicate the best and the second best, respectively.}
    \label{tab:lifelong-prauc-scores}
    \scalebox{0.85}{
    \begin{tabular}{lcccccccc}
        \toprule
        \textbf{Model} & \textbf{X-IIoTID} & \textbf{UNSW-NB15} & \textbf{WUSTL-IIoT} & \textbf{MQTTset} & \textbf{CIC-IDS 2017} & \textbf{CIC-IDS 2018} & \textbf{Average} & \textbf{Std. Dev.} \\
        \midrule
        \textbf{CITADEL} & 0.6596 & 0.8045 & 0.9785 & 0.7274 & 0.5449 & 0.4745 & 0.7060 & 0.1811 \\
        \textbf{ADCN} & \textbf{0.8520} & 0.7287 & 0.9706 & 0.6295 & \textbf{0.7085} & \underline{0.4781} & \underline{0.7279} & \textbf{0.1627} \\
        \textbf{VLAD} & 0.5755 & 0.7314 & 0.9593 & \textbf{0.8428} & 0.4967 & 0.3144 & 0.6534 & 0.2242 \\
        \textbf{SSF} & 0.4935 & 0.7349 & 0.9591 & \underline{0.8026} & 0.2370 & 0.3399 & 0.5945 & 0.2577 \\ 
        \textbf{LOF} & \underline{0.7077} & 0.9050 & \underline{0.9972} & 0.6998 & \underline{0.6227} & \textbf{0.5082} & \textbf{0.7401} & \underline{0.1737} \\
        \textbf{IF} & 0.6131 & \textbf{0.9284} & \textbf{0.9993} & 0.7591 & 0.4001 & 0.3173 & 0.6696 & 0.2571 \\
        \textbf{SGDOCSVM} & 0.6297 & \underline{0.9097} & 0.9961 & 0.4945 & 0.4031 & 0.3015 & 0.6224 & 0.2670 \\
        \textbf{SLAD} & 0.3899 & 0.8441 & 0.9615 & 0.3205 & 0.3800 & 0.3839 & 0.5467 & 0.2730 \\
        \textbf{ICL} & 0.3992 & 0.8404 & 0.9615 & 0.3692 & 0.3359 & 0.3259 & 0.5387 & 0.2470 \\
        \textbf{RCA} & 0.3729 & 0.8479 & 0.9859 & 0.2203 & 0.2012 & 0.3452 & 0.4956 & 0.2871 \\
        \textbf{RDP} & 0.3634 & 0.8546 & 0.9854 & 0.2983 & 0.2174 & 0.2427 & 0.4936 & 0.2892 \\
        \textbf{Anomal-E} & 0.5488 & 0.7537 & 0.9543 & 0.7129 & 0.5402 & 0.4418 & 0.6586 & 0.1804 \\
        \bottomrule
    \end{tabular}}
\end{table*}

\begin{table*}[t]
    \centering
    \caption{SOTA BWT Comparison. Bold and underlined cases indicate the best and the second best, respectively.}
    \label{tab:bwt-scores}
    \scalebox{0.85}{
    \begin{tabular}{lcccccccc}
        \toprule
        \textbf{Model} & \textbf{X-IIoTID} & \textbf{UNSW-NB15} & \textbf{WUSTL-IIoT} & \textbf{MQTTset} & \textbf{CIC-IDS 2017} & \textbf{CIC-IDS 2018} & \textbf{Average} & \textbf{Std. Dev.} \\
        \midrule
        \textbf{CITADEL} & \textbf{0.1414} & \textbf{0.1523} & \textbf{0.3150} & 0.1674 & 0.2450 & 0.2247 & \textbf{0.2076} & \underline{0.0609} \\
        \textbf{ADCN} & 0.0021 & -0.0084 & -0.0050 & -0.0057 & -0.0118 & -0.0212 & -0.0083 & \textbf{0.0078} \\
        \textbf{VLAD} & -0.2161 & -0.3154 & -0.0248 & -0.1574 & -0.4840 & -0.3148 & -0.2521 & 0.1531 \\
        \textbf{SSF} & 0.0410 & 0.0085 & \underline{0.0033} & 0.0531 & -0.3367 & -0.1245 & -0.0592 & 0.1366 \\
        \textbf{LOF} & -0.1496 & -0.0449 & -0.0026 & -0.2202 & -0.4466 & -0.3835 & -0.2079 & 0.1713 \\
        \textbf{IF} & -0.1608 & -0.0489 & -0.0781 & -0.2707 & -0.3759 & -0.2059 & -0.1901 & 0.1050 \\
        \textbf{SGDOCSVM} & -0.1524 & -0.0220 & -0.0013 & -0.2165 & -0.3705 & -0.3399 & -0.1838 & 0.1373 \\
        \textbf{SLAD} & 0.1049 & 0.0512 & 0.0004 & 0.0092 & \textbf{0.3903} & \textbf{0.3428} & 0.1498 & 0.1487 \\
        \textbf{ICL} & \underline{0.1196} & \underline{0.0535} & 0.0004 & \textbf{0.4489} & \underline{0.3243} & \underline{0.2822} & \underline{0.2048} & 0.0644 \\
        \textbf{RCA} & 0.0606 & 0.0552 & -0.0003 & 0.3118 & 0.1200 & 0.2398 & 0.1312 & 0.1178 \\
        \textbf{RDP} & 0.0624 & 0.0463 & 0.0004 & \underline{0.3192} & 0.0000 & 0.0000 & 0.0714 & 0.1235 \\
        \textbf{Anomal-E} & -0.1317 & -0.0805 & -0.0184 & -0.1220 & -0.1618 & -0.1493 & -0.1106 & 0.0431 \\
        \bottomrule
    \end{tabular}}
\end{table*}

\begin{table*}[t]
    \centering
    \caption{SOTA FWT Comparison. Bold and underlined cases indicate the best and the second best, respectively.}
    \label{tab:fwt-scores}
    \scalebox{0.85}{
    \begin{tabular}{lcccccccc}
        \toprule
        \textbf{Model} & \textbf{X-IIoTID} & \textbf{UNSW-NB15} & \textbf{WUSTL-IIoT} & \textbf{MQTTset} & \textbf{CIC-IDS 2017} & \textbf{CIC-IDS 2018} & \textbf{Average} & \textbf{Std. Dev.} \\
        \midrule
        \textbf{CITADEL} & 0.6746 & \textbf{0.9186} & 0.9557 & 0.7025 & \textbf{0.6740} & 0.2777 & \underline{0.7005} & \underline{0.2209} \\
        \textbf{ADCN} & \textbf{0.8436} & 0.6807 & 0.9692 & 0.6285 & 0.2012 & 0.0270 & 0.5584 & 0.3528 \\
        \textbf{VLAD} & 0.5325 & 0.6864 & 0.9385 & \textbf{1.0000} & 0.5908 & 0.2530 & 0.6669 & 0.2735 \\
        \textbf{SSF} & 0.4857 & 0.7977 & 0.9463 & 0.8309 & 0.1618 & 0.3570 & 0.5966 & 0.2817 \\
        \textbf{LOF} & 0.3201 & \underline{0.9071} & \textbf{0.9892} & 0.7096 & 0.2055 & \underline{0.4233} & 0.5925 & 0.2933 \\
        \textbf{IF} & \underline{0.6860} & 0.8584 & \underline{0.9752} & \underline{0.8041} & \underline{0.6005} & 0.1623 & 0.6811 & 0.2866 \\
        \textbf{SGDOCSVM} & 0.5503 & 0.9061 & 0.9565 & 0.5428 & 0.2392 & \textbf{0.4582} & 0.6089 & 0.2473 \\
        \textbf{SLAD} & 0.4868 & 0.8338 & 0.9283 & 0.0301 & 0.4054 & 0.3374 & 0.5036 & 0.3266 \\
        \textbf{ICL} & 0.5010 & 0.8210 & 0.9329 & 0.1703 & 0.4142 & 0.1842 & 0.5039 & 0.2901 \\
        \textbf{RCA} & 0.3960 & 0.8112 & 0.9160 & 0.0212 & 0.3519 & 0.2958 & 0.4654 & 0.3164 \\
        \textbf{RDP} & 0.3210 & 0.8339 & 0.9160 & 0.2482 & 0.2673 & 0.1880 & 0.4624 & 0.2662 \\
        \textbf{Anomal-E} & 0.5176 & 0.8204 & 0.9535 & 0.6944 & 0.9509 & 0.4751 & \textbf{0.7353} & \textbf{0.1866} \\
        \bottomrule
    \end{tabular}}    
\end{table*}

\begin{figure}[t]
    \centering
    \captionsetup{justification=centering}
    \includegraphics[width=1\linewidth]{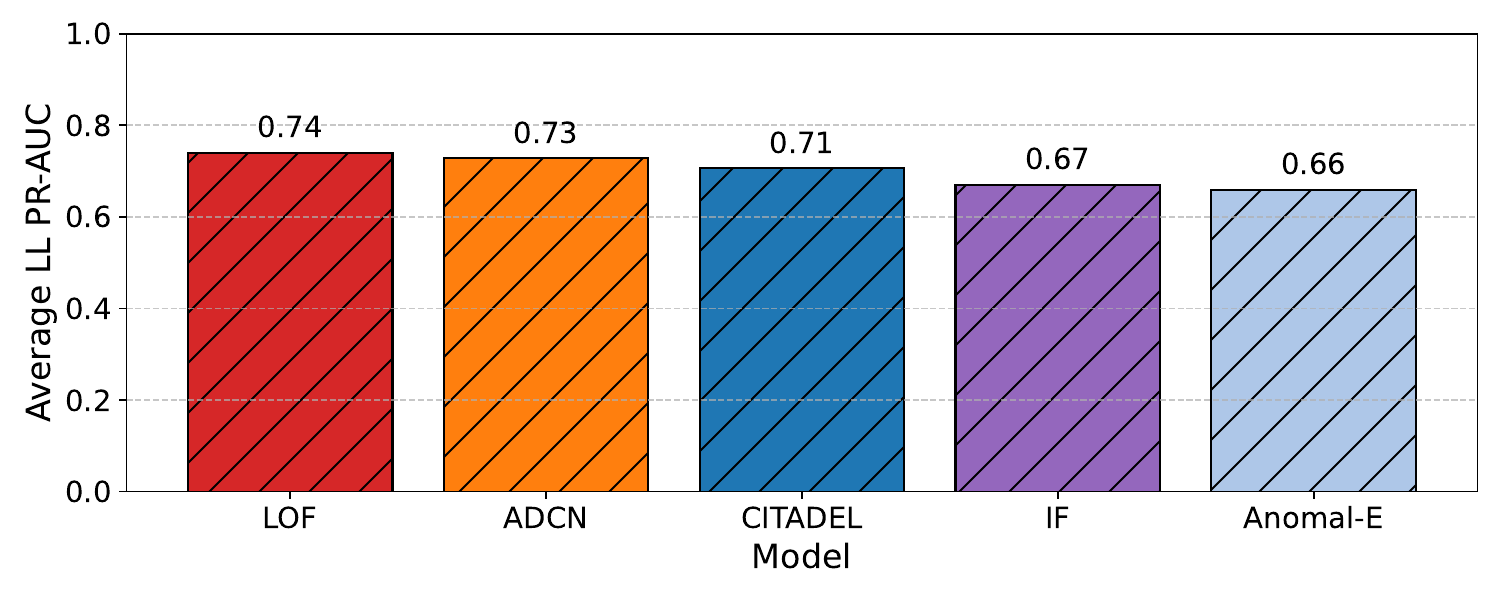}
    \includegraphics[width=1\linewidth]{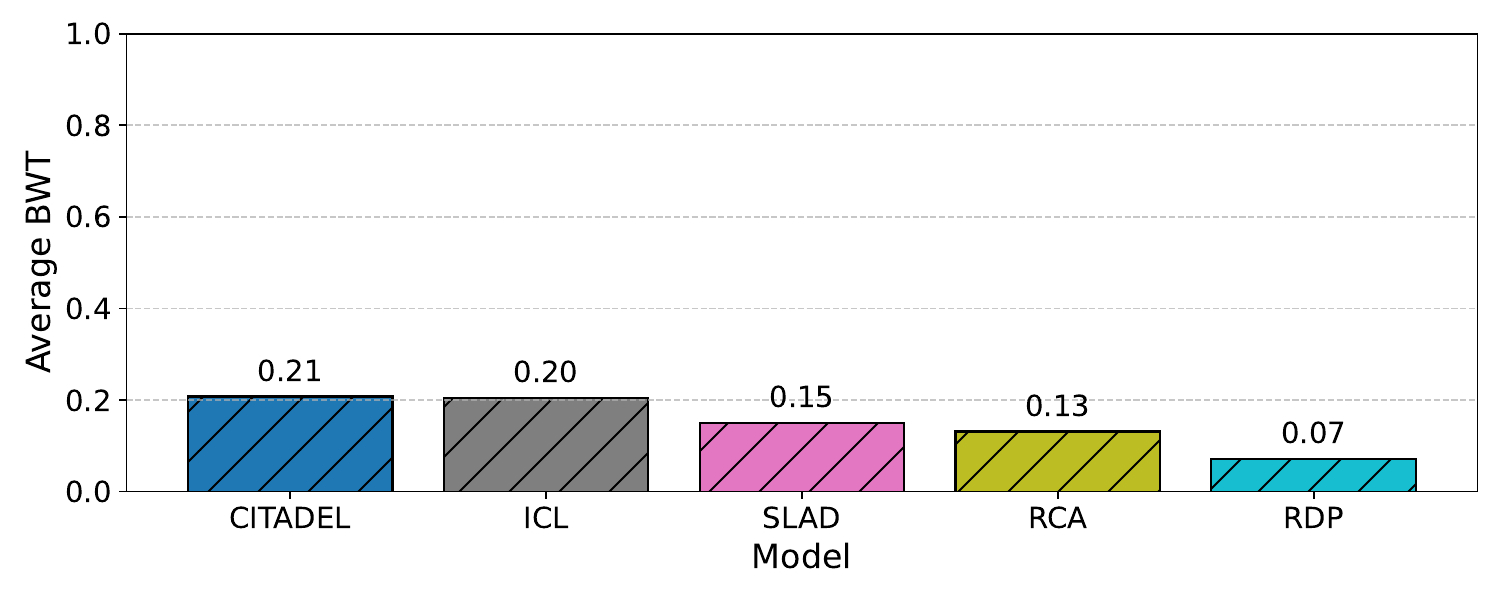}
    \includegraphics[width=1\linewidth]{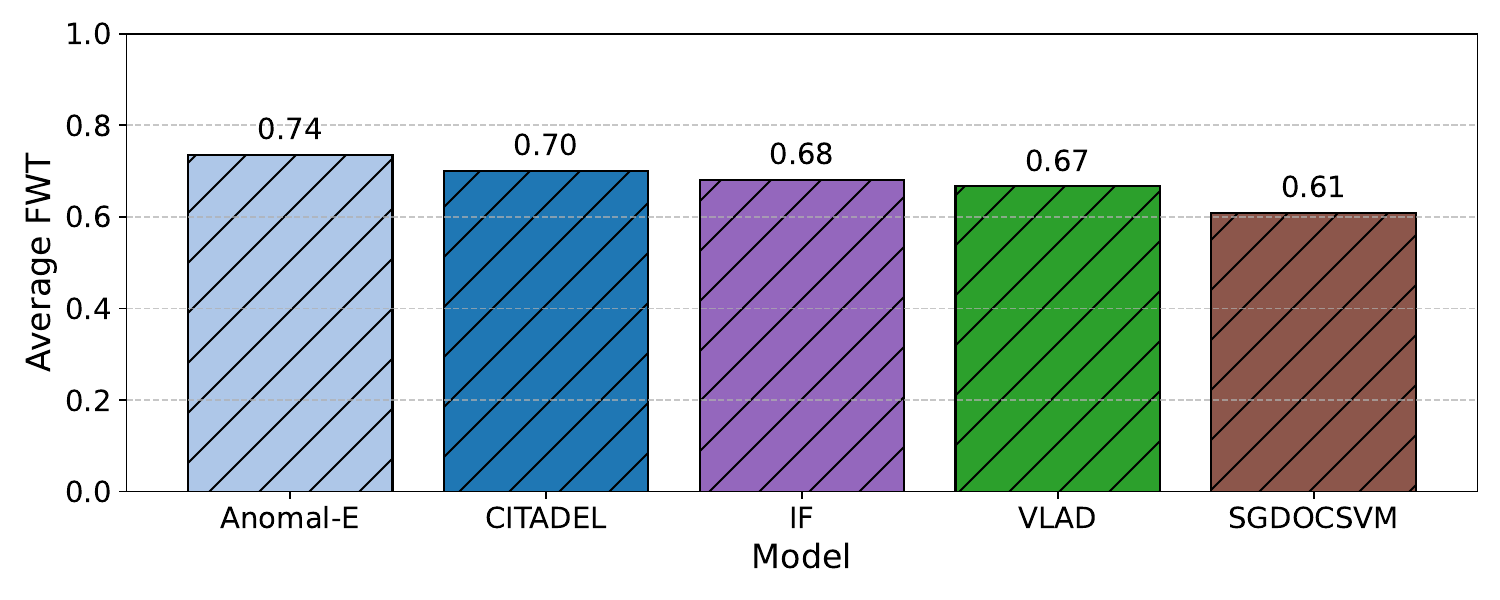}
    \caption{Top 5 models averaged across datasets based on (top) Lifelong PR-AUC, (middle) Backward Transfer (BWT), and (bottom) Forward Transfer (FWT).}
    \label{fig:top5-stacked}
\end{figure}

\paragraph{CITADEL Results}
Tables~\ref{tab:lifelong-prauc-scores}, \ref{tab:bwt-scores}, and \ref{tab:fwt-scores} present the LL PR-AUC, BWT, and FWT scores, respectively.

\textbf{LL PR-AUC: }As shown in Table \ref{tab:lifelong-prauc-scores}, LL PR-AUC scores were computed for each model on a per-dataset basis, followed by average and standard deviation aggregation. CITADEL yields a strong average LL-PR-AUC at 70.6\%, which ranks within the top three, with a particularly solid performance on WUSTL-IIoT (97.85\%), indicating effective precision-recall tradeoffs across a diverse set of environments. CITADEL maintains a relatively stable performance with a standard deviation of 18.1\%, suggesting robust generalization across different distributions.

\textbf{BWT: }Table \ref{tab:bwt-scores} shows that CITADEL exhibits positive BWT on all six datasets, with the highest scores on X-IIoTID (14.14\%), WUSTL-IIoT (31.50\%) and UNSW-NB15 (15.23\%), indicating strong knowledge consolidation and resistance to catastrophic forgetting. The model’s average BWT of 20.76\% is the highest among all models, closely followed by ICL at 20.48\%. In contrast, most baseline models, including VLAD, LOF, and IF, exhibit consistently negative BWT, confirming substantial performance degradation on prior tasks. Compared to SOTA methodologies, CITADEL has improvements of up to 32\% against ADCN (WUSTL-IIoT) and gains up to 72.9\% compared to VLAD (CIC-IDS 2017).

\textbf{FWT: }As detailed in Table \ref{tab:fwt-scores}, Anomal-E slightly outperforms CITADEL with an average FWT of 73.53\%, while CITADEL closely follows with 70.05\%. However, CITADEL demonstrates more consistent generalization across datasets, achieving the highest FWT on UNSW-NB15 (91.86\%) and CIC-IDS 2017 (67.4\%). Compared to models like ADCN and LOF, whose average FWTs fall below 60\%, CITADEL exhibits stronger forward knowledge transfer with a lower standard deviation (22.09\%), indicating stable learning across tasks. These consistent gains reinforce the design of CITADEL, which enables proactive adaptation in evolving network environments while maintaining past knowledge.

\textbf{Average Performance: }Figure \ref{fig:top5-stacked} visualizes the top five models in terms of average performance across three continual learning metrics (LL PR-AUC, BWT, and FWT). Each bar plot aggregates results across six benchmark datasets, offering a holistic view of generalization, retention, and transfer capabilities. Notably, CITADEL consistently ranks among the top-performing models across all three metrics, achieving an average LL-PR-AUC of 70.60\%, a BWT of 20.76\%, and an FWT of 70.05\%. These results highlight CITADEL's ability to balance forward generalization with backward retention.

\textbf{Insights:} CITADEL effectively mitigates catastrophic forgetting while adapting to new tasks, a key requirement for intrusion detection systems operating in dynamic threat environments. Compared to prior methods, CITADEL achieves a strong balance between retaining knowledge of earlier attacks and incorporating new information, indicating its suitability for continual learning in IoT security contexts.

\begin{table}[t]
    \centering
    \caption{CITADEL Ablation Study (X-IIoTID)}
    \label{tab:citadel-strategy-comparison}
    \scalebox{0.9} {
    \begin{tabular}{lccc}
        \toprule
        \textbf{Strategy} & \textbf{LL PR-AUC} & \textbf{BWT} & \textbf{FWT} \\
        \midrule
        CITADEL & \textbf{0.6596} & \textbf{0.1414} & \textbf{0.6746} \\
        Only Strategic Sampling and Forgetting & 0.4935 & 0.0410 & 0.4857 \\
        Only Hierarchical Memory & 0.4261 & 0.0013 & 0.5246 \\
        \bottomrule
    \end{tabular}}
\end{table}

\paragraph{Ablation Study} 
To further investigate the contributions of individual components within CITADEL, we conducted an ablation study by systematically removing each component and evaluating the resulting impact on performance. Since our method contains two main components, strategic sampling/forgetting and a hierarchical memory, we experimented with removing the hierarchical memory and directly optimizing the memory buffer as well as storing new data into the hierarchical memory without any KL-divergence optimization process. Table \ref{tab:citadel-strategy-comparison} presents our evaluation on the X-IIoT dataset, reporting results for three metrics: LL PR-AUC, BWT, and FWT. Comparing the performances, we can see that when comparing every metric, CITADEL performs significantly better with scores  of 65.96\% (+16.61\% from Strategic Sampling and Forgetting), 14.14\% (+10.04\% from Strategic Sampling and Forgetting), and 67.46\% (+15.00\% from Hierarchical Memory). Due to these results, the inclusion of both components within CITADEL is crucial towards maximizing performance in a lifelong setting.

\begin{table}[t]
    \centering
    \small
    \caption{Inference Time per Sample (X-IIoTID)}
    \label{tab:overhead-analysis-citadel}
    \scalebox{0.80} {
    \begin{tabular}{lcccccc}
        \toprule
        \textbf{Detector} & \textbf{CITADEL} & \textbf{SSF} &\textbf{RDP} & \textbf{RCA} & \textbf{ICL} & \textbf{SLAD} \\
        \midrule
        \textbf{Time (ms)} & 0.0493 & 0.0524 & 0.0511 & 0.1351 & 0.0481 & 0.3795 \\
        \bottomrule
    \end{tabular}}
\end{table}

\paragraph{Overhead Analysis}
Computational overhead of ML models in IoT IDS is critical, as resource-constrained IoT devices require real-time threat detection without significant delays to ensure prompt security responses. 
In this context, Table \ref{tab:overhead-analysis-citadel} presents the inference time per sample overhead analysis for the top six best performing methods on the X-IIoTID dataset in terms of BWT, measured in milliseconds. CITADEL achieves significantly more efficient inference speed compared to all competing models, with the exception of ICL, which offers only a negligible speed advantage of 0.0012 milliseconds per sample. Importantly, CITADEL consistently surpasses ICL across all three CL metrics, demonstrating its superior overall performance while maintaining near-optimal efficiency. Furthermore, given that the median traffic flow has a duration of $4.98$ milliseconds~\cite{gungor2024roldef}, the overhead introduced in CITADEL's overhead is a negligible proportion of only $0.98\%$.

\section{Conclusion}\label{sec:conclusions}
The rise of IoT devices has intensified the need for machine learning-based intrusion detection systems (ML-IDS) to secure increasingly dynamic and heterogeneous networks. However, ML-IDS face critical challenges, including dependency on labeled attack data and limited adaptability to zero-day threats. Self-supervised learning (SSL) addresses these gaps by enabling models to learn patterns from unlabeled data, enhancing their capacity to identify emerging threats. The addition of a continual learning (CL) methodology is vital to ensure that the models are functional in real-world scenarios where network intrusion data comes in various forms, fostering adaptation to new types of attacks. To this end, we propose CITADEL, a novel end-to-end framework that combines self-supervised and continual learning. CITADEL restructures tabular data into an image-like representation, enabling Masked Autoencoders (MAEs) to extract rich insights into network behavior. Together with our integrated KL-divergence optimization and hierarchical memory buffer, this design yields a robust and adaptive system that effectively retains previously learned knowledge while efficiently incorporating new information. Experimental results across six benchmark intrusion detection datasets demonstrate that CITADEL not only achieves state-of-the-art performance in zero-day attack detection but also maintains strong generalization and transferability in lifelong learning settings, with BWT improvement up to 72.9\% over VLAD in key detection and retention metrics.

\section*{Acknowledgments}
This work has been funded in part by NSF, with award numbers \#1826967, \#1911095, \#2003279, \#2052809, \#2100237, \#2112167, \#2112665, and in part by PRISM and CoCoSys, centers in JUMP 2.0, an SRC program sponsored by DARPA.

\bibliographystyle{IEEEtran}
\bibliography{biblio}

\end{document}